\documentclass[%
reprint,
 amsmath,amssymb,
 aps,
prstab,
floatfix,
]{revtex4-1}

\usepackage{amsmath}
\usepackage{amssymb}
\usepackage{latexsym}
\usepackage{graphicx}% Include figure files
\usepackage{dcolumn}% Align table columns on decimal point
\usepackage{bm}% bold math
\begin{document}

\title{Drop impact on superheated surfaces}

%\author{Tuan Tran}
%\email{t.tran@utwente.nl}
%\author{Erik-Jan Staat}
%\author{Chao Sun}
%\email{c.sun@utwente.nl}
%\author{Andrea Prosperetti}
%\altaffiliation[Also at ]{Department of Mechanical Engineering, Johns Hopkins University, Baltimore, MD 21218, USA}
%\author{Detlef Lohse}
%\affiliation{Physics of Fluids, University of Twente, P.O. Box 217, 7500 AE Enschede, The Netherlands}

\author{Tuan Tran$^1$}
\email{t.tran@utwente.nl}
\author{Hendrik J. J. Staat$^1$}
\author{Andrea Prosperetti$^{1,2}$}
\author{Chao Sun$^1$}
\email{c.sun@utwente.nl}
%\altaffiliation[Also at ]{Department of Mechanical Engineering, Johns Hopkins University, Baltimore, MD 21218, USA}
\author{Detlef Lohse$^1$}
%\affiliation{Physics of Fluids, University of Twente, P.O. Box 217, 7500 AE Enschede, The Netherlands}
\email{d.lohse@utwente.nl }
\affiliation{
$^1$Physics of Fluids, University of Twente, P.O. Box 217, 7500 AE Enschede, The Netherlands\\
$^2$Department of Mechanical Engineering, Johns Hopkins University, Baltimore, MD 21218, USA
}

\date{\today}% It is always \today, today,
             %  but any date may be explicitly specified

\def\Re{\¤{Re}}
\def\We{\textrm{We}}

\begin{abstract}
At impact of a liquid droplet on a smooth surface 
heated above the liquid's boiling point, 
the droplet either immediately
boils when it contacts the surfaces (``contact boiling''), 
or without any surface contact forms a Leidenfrost vapor layer
towards the hot surface and  bounces back (``gentle film boiling''), 
or 
both forms the Leidenfrost layer and ejects tiny droplets upward
(``spraying film boiling'').  
We experimentally determine conditions under which 
impact behaviors in each regime can be 
realized.
We show that the dimensionless maximum spreading $\gamma$ of 
impacting droplets on the heated surfaces in both 
gentle and spraying film boiling regimes 
shows a universal scaling with the Weber number $\We$ ($\gamma\sim\We^{2/5}$) -- regardless 
of surface temperature and of liquid properties -- which is much 
steeper than for the impact on non-heated (hydrophilic or hydrophobic) surfaces  ($\gamma\sim\We^{1/4}$). 
We also intereferometrically measure the vapor thickness under the droplet.
\end{abstract}

\maketitle

%%%%%%%%%%%%%%%%%%%%%%%%%%%%%%%%%
%%%%%Introduction
When a drop 
impinges gently 
%is gently deposited
on a surface heated well above the 
liquid's boiling temperature, 
%evaporation temperature,
the liquid 
may evaporate so fast 
that the drop floats on its own vapor. 
The vapor layer then acts as a thermally
insulating film causing the drop to 
evaporate much more slowly than 
if it remained 
in contact with the surface.
This phenomenon is known as
the Leidenfrost effect \cite{leiden1756}; 
the temperature
at which the evaporation time of the drop 
reaches its maximum is called
the Leidenfrost temperature. 
%This definition of the Leidenfrost 
%temperature neglects
%the impact dynamics, which 
%By definition, the Leidenfrost temperature
%is independent of the impact dynamics
%At a critical temperature, known 
%as the Leidenfrost temperature ,
%the evaporation time of the drop 
%reaches its maximum, and 
%correspondingly the heat flux 
%is minimum. 
Since it was first reported in 1756,
%the Leidenfrost effect has been the
%subject of intense investigations
%due to its particular importance in many
%industrial processes 
%\cite{FILL}.
various aspects of the Leidenfrost effect
have been studied, %among which, 
most importantly the determination of the 
Leidenfrost temperature for different liquids and surfaces
\cite{bernardin99,gottfried66}.
%Most of these studies have 
%%relied on
%%the assumption of negligible impact velocity to 
%assumed negligible impact velocity in order to
%estimate the Leidenfrost temperature
In general, measurements of the Leidenfrost temperature 
were performed with zero or at most small incident velocity because the
characteristic time scale of the impact,
%generally 
of order of several milliseconds,
is negligible compared to the
drop's total evaporation time. 
In other words, the Leidenfrost temperature
is assumed not to be affected by the impact dynamics
(hence to be referred herein as the 
\emph{static Leidenfrost temperature}), 
and is 
commonly considered as the lowest boundary
of the film boiling regime
\cite{bernardin97,chandra91,anders93,xiong91}.
%We refer to this as the static Leidenfrost temperature. 
However, in most realistic situations where the impact 
velocity is not negligible, the Leidenfrost temperature
should be regarded as a dynamic quantity 
%\cite{gottfried66} (see review article \cite{moreira10}) at which the vapor layer
%bounces an impinging droplet. 
%In that case, 
%%in contrast to an extensive body of work 
%%on the static Leidenfrost temperature, 
%no definite conclusion can be made as to
%what the dynamic Leidenfrost temperature would be
%and how it depends on the impact conditions.
\cite{gottfried66} (see review articles \cite{moreira10,rein02}).
%Here we define the \emph{dynamic Leidenfrost temperature}
%as the surface's minimum temperature at which the 
%developing vapor layer causes an impinging droplet
%to bounce. 
%In that case, 
%%in contrast to an extensive body of work 
%%on the static Leidenfrost temperature, 
%no definite conclusion can be made as to
%what the dynamic Leidenfrost temperature would be
%and how it depends on the impact conditions.
One can define the \emph{dynamic Leidenfrost temperature}
$T_L$ as the minimum temperature of the surface at which the 
developing vapor layer causes an impinging droplet
to bounce. As compared to the static case, there have been
very few studies that focus on the dependence of $T_L$ on
 impact conditions.
%Then a prediction can be made that $T_L$ depends on
%the impact conditions. 
The goal of this Letter is to experimentally 
determine this dependence and to study 
 droplet impact dynamics on heated surfaces.

%%%%% Experimental setup
% accompanied by figure showing experimental setup
\begin{figure}[h!]
\begin{center}
\includegraphics[width=5.4cm]{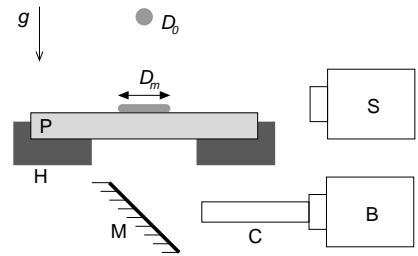}
%\spacing{1}
\caption{\small{ 
Schematic of the experimental setup (not drawn to scale) 
used to study droplet impact on heated surfaces.
A liquid droplet of initial diameter $D_0\approx 2\,$mm falls on a heated
plate P and spreads to its maximum diameter $D_m$. 
The plate P is a polished silicon plate 
(average roughness $\approx5\,$nm) 
in most experiments.
In the case that a bottom view is needed, a polished sapphire plate is used 
instead of the silicon one. The plate is placed on a holder H which 
can be heated by six cartridge heaters
embedded symmetrically inside. The heaters are monitored by
a controller via a solid-state switch (Omega Inc.).
The maximum temperature that can be obtained with this 
system is $700^o$C, with accuracy within 1K. 
The holder has a $2\,$cm-diameter 
hole in the center allowing bottom-view observation.
The side view of the impact is recorded by camera S (Photron SA1), and 
the bottom view is recorded by camera B (Photron SA2) connected 
to a long working distance microscope C via a mirror M. 
%The two high-speed cameras are Photron Fastcams SA1 $\&$ SA2.
}}
\label{fig:setup}
\end{center}
\end{figure}
For this purpose, 
%we let droplets fall on a heated smooth surface. 
%that can be heated up to $800^o$C.
we generate droplets
by pushing liquid from a syringe 
at a small rate ($\approx 0.05\,$ml/min)
through a pipe and into a capillary needle (inner diameter $0.1\,$mm).
%The flow rate is kept very small to make
%sure that the drop falls only under the
%gravitational effect.
The droplet 
formed 
at the needle's tip 
detaches as soon as the gravitational force
overcomes the surface tension.
%Thus, the drop size is %accurately 
%highly reproducible for each choice of liquid 
%and needle. 
We use two different liquids: milli-Q water
(density $\rho_w = 998\,$kg/m$^3$,
surface tension $\sigma_w = 72\times 10^{-3}\,$N/m,
viscosity $\nu_w = 10^{-6}\,$m$^2$/s),
and FC-72 
($\rho_{fc} = 1680\,$kg/m$^3$,
$\sigma_{fc} = 10\times 10^{-3}\,$N/m,
$\nu_{fc} = 0.38\times10^{-6}\,$m$^2$/s).
%let liquid droplets fall under gravity 
%from a capillary needle at a distance $H$ fromthe surface. 
%The drop size is accurately reproducible
%and is determined by the balance between 
%gravity and surface tension that hangs the 
%drop.
%The flow rate at which the liquid is pushed 
%out from the needle is very small 
%in order to 
%The liquid is pushed out of the needle
%by a syringe pump at a constant speed ($\approx 1\,$ml/min)
%to ensure the drop's reproducibility. 
By varying the 
%distance $H$ over which the droplet
%accelerates under gravity, 
needle's height,
we control the velocity $V$ of the droplet before
impacting the surface. 
We simultaneously capture
side-view and bottom-view images
of the droplet as it spreads using
two synchronous high-speed cameras 
(Photron Fastcam SA1 $\&$ SA2). 
From the series of recorded 
images in each experiment, 
we obtain the impact velocity $V$,
the drop diameter $D_0$ 
(typically $1.7\,$mm and $2.2\,$mm for water and 
$1.1\,$mm for FC-72),
and the maximum spreading 
diameter $D_m$ (Fig.~\ref{fig:setup}).
As a result, 
we can estimate
the drop's kinetic energy
compared to its surface energy 
by computing
the Weber number 
$\textrm{We} = \rho D_0V^2/\sigma$.
%This dimensionless number is then used
%as a control parameter in our
%experiments.
In our experiments, We is varied from 
0.5 to 500 for water and from 6 to 600 for FC-72.

%Surface, how to heat, what kinds, how to accommodate bottom view
%Argument about surface's thermal diffusivity and impact duration
%The heated surfaces are silicon wafer 
%and sapphire plate.
%In most of our experiments, 
%we use polished silicon plates 
%as test surfaces. 
%In some cases where bottom-view
%visualization is desired, 
%we use a polished sapphire plate instead of
%the silicon one. For this purpose,
%the holder where the 
The test surfaces in most of 
our experiments are polished 
silicon plates
(silicon wafers, average surface roughness $\approx 5\,$nm).
The plate is placed on top of 
a stainless-steel holder (Fig.~\ref{fig:setup}).
At the center of the holder, 
%we bore a $2\,$cm-diameter hole 
%so that bottom-view observation is 
%possible if a sapphire plate
a $2\,$cm-diameter hole 
allows for bottom-view observations if a sapphire plate 
is used instead of the silicon one.
%This is crucial in
%our study as it allows us to 
%identify characteristic behaviors
%of the drop during impact for each regime.
We embed in the holder a temperature probe and six cartridge heaters
symmetrically around the hole
to control its temperature and 
consequently the temperature of 
the test surface.
Since sapphire and
silicon both have high thermal conductivity, 
the temperature difference between the 
holder and the test surface is 
only a few degrees and can be 
neglected in the explored temperature
range (from $200^o$C to $600^o$C).
As a result, the surface temperature $T$ is 
approximated as the holder's temperature.

%%%%%Characterization of each regime
%accompanied by series of images of drops during impact
% for each regime
%Impact of droplets on heated surfaces differs
%from that of droplets on normal surfaces
%Depending on the Weber number 
%and the surface temperature, 
%experiments of droplet impact may
%have many different outcomes.

\begin{figure}[hpdp]
\begin{center}
\includegraphics[width=6.3cm]{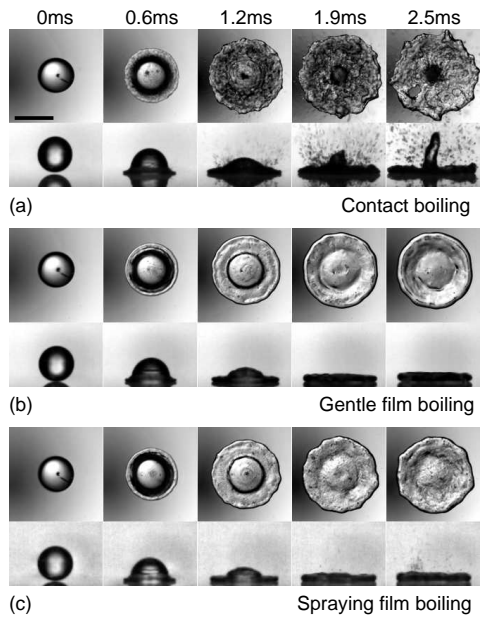}
%\spacing{1}
\caption{\small{ 
Series of images of representative water droplet impacts
in three regimes. The Weber number in all 
three experiments is 32. Each image has both
bottom view and side view of the impact. Images
in the same column are taken at the same time after 
impact. (a)The surface temperature $T=380^o$C; 
Contact boiling.
The first sign of droplet ejection is 
seen at $0.6\,$ms after impact. 
(b) $T=500^o$C; Gentle film boiling.
(c) $T=580^o$C; Spraying film boiling. 
The contrast of the side-view images was enhanced
to show tiny droplets ejected upward at $1.9\,$ms.
The inset bar (shown in upper left image) 
indicates a length scale of $2.5\,$mm.
}}
\label{fig:series}
\end{center}
\end{figure}

%To demonstrate different impact behaviors of droplets, 
%we show in Fig.~\ref{fig:series} 
%series of images taken %during impact of water droplet 
%in three representative experiments with water droplets: each one 
%corresponds to an impact regime.
%%These specific examples are representative of 
%%different impact regimes.  	
%We vary the surface temperature 
%while keeping the Weber number fixed in these 
%experiments in order to 
%compare
%characteristic behaviors between different regimes.
We repeat the droplet impact experiment 
numerous times
using water as the working fluid  
for different Weber numbers ($0.5\leqslant \We \leqslant 500$) and surface
temperatures ($250^o{\rm C} \leqslant T \leqslant 560^o{\rm C}$), and 
observe the drop behavior during impact.
%We categorize behaviors of the drop during impact 
%in different regimes.
% (typical behavior in each 
%regime is shown ).
%are shown
%in Fig.~\ref{fig:series}).
%The phase diagram in 
Figure~\ref{fig:series} shows three
distinct
impact regimes, each one of which is 
exemplified by a series of images 
taken from a high-speed recording of 
a representative experiment. 

\begin{figure}[hpdp]
\begin{center}
\includegraphics[width=7cm]{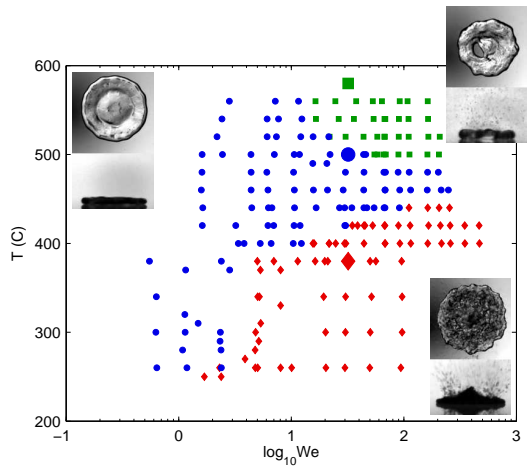}
%\spacing{1}
\caption{\small{ 
Phase diagram for water droplet impact on a heated surface
showing three separate regions: contact boiling regime 
(red solid diamonds), 
gentle film boiling regime (blue solid circles),
spraying film boiling regime (green solid squares). 
Each region has an inset illustrating 
typical droplet impact behavior in that regime. 
The three large symbols represent the experiments shown 
in Fig.~\ref{fig:series}.
}}
\label{fig:phase_diagram}
\end{center}
\end{figure}

Fig.~\ref{fig:series}a shows 
images of an experiment 
%(with $\We= 32$ and $T=380^o$C) 
in the
\emph{contact boiling regime}.
The bottom views in these images evidently show that
shortly after impact, the liquid makes 
partial contact with the surface. The contact leads 
to a high rate of heat transfer from the heated surface 
and consequently
formation and growth of vapor bubbles.
The vapor pressure increases abruptly
causing disruption of the liquid's bottom surface, 
as well as violent, sometimes explosive, 
ejection of tiny droplets due to the venting of the vapor bubbles
(clearly seen from the side views).
In the phase diagram (Fig.~\ref{fig:phase_diagram}),
this regime corresponds to the region in red color (diamonds).

The \emph{gentle film boiling regime}
is shown in Fig.~\ref{fig:series}b. 
The name refers 
to situations in which the vapor layer 
is sufficiently thick to prevent the liquid 
from touching the surface 
and there is no droplet ejection due to 
expansion of vapor bubbles (disintegration of the impacting droplet
may happen, but due to other mechanisms, 
e.g., instability at the rim of the spreading lamella at high
Weber number). 
This regime corresponds to the region in blue color (circles)
in Fig.~\ref{fig:phase_diagram}.
%;snapshots of the experiment showing 
%its typical behavior is presented in .

The \emph{spraying film boiling regime} (Fig.~\ref{fig:series}c)
is similar to the gentle film boiling regime in 
that the liquid is not in contact with 
the surface (bottom views in Fig.~\ref{fig:series}c). 
However, the side-view images
reveal spraying-like ejection of small droplets from the top 
of the liquid, although the ejection is not as vigorous 
as in the case of contact boiling. 
This regime corresponds to the region 
in green color (squares) in Fig.~\ref{fig:phase_diagram}.

%%%%Phase diagram and description
%%%%%Explanation of the phase diagram and transitional trends

%Now we discuss
Let us discuss 
the transition between 
the contact boiling and the gentle film boiling regimes
in Fig.~\ref{fig:phase_diagram}.
This transition %essentially
marks the dependence of the 
dynamic Leidenfrost temperature $T_L$ on the Weber number $\We$. 
%Here, the dynamic Leidenfrost temperature is the minimum 
%point 
%at which the vapor layer can
%suspend an impinging droplet \cite{gottfried66}.
While there
have been disparate conclusions regarding
%how it depends on the Weber number 
%\cite{yao88,bernardin04,wang00,celata06}, 
whether
%it
$T_L$ increases 
\cite{yao88,bernardin04},
or decreases
\cite{wang00, celata06} with $\We$, 
our data show 
%evidently
%unequivocally 
unambiguously
that $T_L$
increases along with $\We$. 
%raises with increasing Weber number. 
We account for this change in $T_L$ by
comparing the pressure in the vapor layer and the 
drop's dynamic pressure: an impinging droplet bounces back from 
the heated surface (hence in the gentle film boiling regime)
if the vapor pressure overcomes the drop's dynamic pressure.
Note that increasing the surface temperature 
raises the vapor pressure and the drop's dynamic pressure
is essentially determined by $\We$. 
Therefore,
a higher surface temperature
is necessary to keep droplet impact at higher
Weber number in the gentle film 
boiling regime.
As a result, we conclude that the dynamic Leidenfrost temperature 
increases with increasing Weber number, consistent with 
our experimental results.

The second transition 
in Fig.~\ref{fig:phase_diagram}
is between
the gentle and spraying film boiling regimes.
From $\We=11$ and $T = 560^o$C,
%it 
the transition temperature
decreases 
with increasing Weber number. 
To understand this result, 
%To account for this transition,
we argue that droplet ejection
in the spraying film boiling regime 
is caused by the bursting of
vapor bubbles in the liquid film.
%when their vapor pressure overcomes the 
%hydrostatic pressure $P_h$.
As $\We$ is increased, the liquid film 
gets thinner and it is easier for the boiling bubbles
to burst through the liquid's upper surface. 
As a result, the transition temperature decreases 
 as the Weber number increases, in accordance with 
 our experimental results.
We stress that 
the bursting of vapor bubbles in the liquid film 
is a crucial condition for the transition from gentle to spraying 
film boiling regime. We confirm this 
by noting that for a fixed Weber number,
 adding $50\,\mu$m particles to
the liquid effectively reduces the transition temperature
\cite{particles}.  
This observation implies that the transition 
from gentle to spraying film boiling regimes
is related to the vapor bubble formation inside the liquid film,
which is enhanced due to increasing of nucleation sites 
provided by the particles.

%Here the inertial pressure is neglected 
%because droplet ejection 
%for impact close to the transition
%only starts when the drop has reached 
%its maximum deformation $D_m$, at which point 
%the fluid stops moving.
%We verify this by %choosing experiments 
%measuring 
%the time $\tau_e$ to observe first ejection
%for impact in the spraying boiling regime, but close
%to the transitional line, 
%and compare it to the time $\tau_m$ 
%the drop needs to reach $D_m$.
%Typically in our experiments, $\tau_e\approx 2.2\,$ms
%and $\tau_m\approx2.5\,$ms, confirming our assumption.
%We then estimate
%$P_h\approx \rho g h$,
%where $h$ is the thickness of the liquid;
%increasing the Weber number 
%makes the liquid layer thinner and 
%subsequently decreases $P_h$.
%Thus, 
%%to suppress the vapor pressure of the bubbles %bubbles's vapor pressure
%to suppress the bubbles's vapor pressure
%at higher Weber number,
%it is required that the surface temperature is reduced.
%%Thus, at higher Weber number, the bubbles's 
%%vapor pressure can break through the liquid sheet
%%at its maximal extension at lower temperature. 
%As a result, we infer that the transition temperature decreases 
% as the Weber number increases, in accordance with 
% our experimental results.

\begin{figure}[hpdp]
\begin{center}
\includegraphics[width=7.5cm]{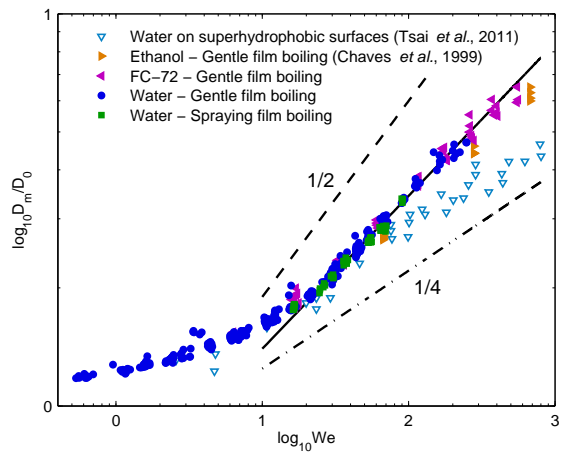}
%\spacing{1}
\caption{\small{ 
Log-log plot of the maximum spreading diameter
normalized by the drop's diameter $(D_m/D_0)$ versus the Weber 
number $(\We)$ for impact in both gentle and spraying film boiling regimes. 
Experimental data for 
water drops spreading on superhydrophobic surfaces at room temperature by 
Tsai {\it et al.} \cite{tsai11} (open downward triangles),
ethanol in the gentle film boiling regime by
Chaves {\it et al.} \cite{chaves99} (solid right triangles), 
FC-72 in the gentle film boiling regime (solid left triangles),
water in the gentle film boiling regime (solid circles),
water in the spraying boiling regime (solid squares).
The Weber number $0.5\leqslant \We \leqslant 600$. 
The surface temperature $250^o$C $\leqslant T \leqslant$ $560^o$C.   
The solid line represents the 
best fit for the experimental data for $\We>10$ 
in the present study with 
the slope 0.39. 
The dashed line represents the scaling $D_m/D_0\sim \We^{1/2}$
resulting from the balance between the drop's initial kinetic
energy and its surface energy at maximum deformation. 
The dashed-dotted 
line represents the scaling $D_m/D_0\sim \We^{1/4}$ 
resulting from a momentum argument \cite{clanet04}.
}}
\label{fig:spread}
\end{center}
\end{figure}

%We stress that 
To obtain a quantitative understanding of 
the spreading dynamics of drop impact on 
the Leidenfrost vapor layer, we measure the 
maximum spreading diameter $D_m$ of the drop
in the gentle and spraying film boiling regimes. 
%(we do not measure the maximum spreading in 
%the contact boiling regime because
%the droplet is mostly boiled violently making 
%it difficult to have reproducible results).
In Fig.~\ref{fig:spread},
we show a log-log plot of the dimensionless maximum spreading
$\gamma=D_m/D_0$ versus the 
Weber number. The plot consists of five sets 
of data: one set was taken using 
water on superhydrophobic surfaces at room temperature
(data by Tsai {\it et al.}, 2011 \cite{tsai11}),
two sets were taken using water 
and FC-72 in the gentle film boiling regime, 
one using water in the spraying boiling regime,
one using ethanol in the gentle film boiling regime 
(data by Chaves {\it et al.}, 1999 \cite{chaves99}). 
Despite a wide variation in surface
temperature ($250^o {\rm C} \leqslant T \leqslant 560^o {\rm C}$),
and differences in liquid (viscosity, surface tension, density) 
and thermal
properties (heat capacity and latent heat of evaporation) 
between water, FC-72, and ethanol, all the data in the gentle and spraying film 
boiling regimes fall on a 
unique, single curve, signaling universality of the 
spreading dynamics in the film boiling regime. 
%We stress that even though the spreading
%of droplets on a Leidenfrost layer is reminiscent of  
%the same process on a superhydrophobic surface,
%the dynamics is essentially different.
For $\We>10$, our data is best fitted by the scaling 
$D_m/D_0 \sim \We^{0.39}\approx\We^{2/5}$.
This is much steeper than the well established 
scaling law $D_m/D_0\sim We^{1/4}$ 
 \cite{clanet04}
found  for the impact of various different liquid droplets on both hydrophilic
 \cite{clanet04}, 
 hydrophobic, and even superhydrophic surfaces
(see \cite{tsai11} and the data of that paper
which we have included in Fig.~\ref{fig:spread}). 
  In this last situation, the liquid spreading is lubricated by 
  an air layer between the drop and the solid surface. 
  Given the universality of the $1/4-$scaling law 
  and the slip due to the air lubrication layer, 
  dissipation clearly does not play a role for the $1/4-$scaling law. 
  The steeper and also universal $0.39-$scaling
   is therefore the more remarkable. 
  This effect may be due to an extra driving mechanism 
   caused by the evaporating vapor radially 
   shooting outwards and taking the liquid along. 
   This interpretation is consistent with the 
   experimentally found ambient pressure 
   dependence of $D_m/D_0$ 
  \cite{tsai11}. 
   Note that balancing the surface energy 
   $\sigma D_m^2$ and the initial kinetic energy 
   $\rho D_0^3 V^2$ would lead to an 
   even steeper scaling $D_m/D_0 \sim We^{1/2}$ 
   which is not observed. 

\begin{figure}[h!]
\begin{center}
\includegraphics[width=5.5cm]{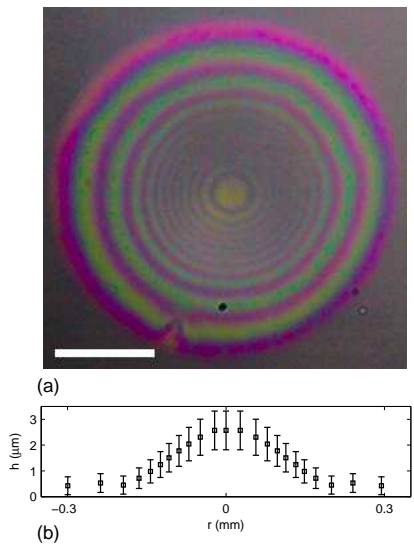}
%\spacing{1}
\caption{\small{(a) Interference pattern showing 
thickness variation of the vapor layer during impact of 
a droplet. The image was 
taken from the bottom view of droplet impact  
using a color high-speed camera connected
to a long working distance microscope at 10000 frame
per second. The Weber number is 3.5 and the surface 
temperature is 350$^o$C. The inset bar indicates 0.2mm.
(b) Profile of the vapor thickness extracted from 
the color image. 
}}
\label{fig:thickness}
\end{center}
\end{figure}

While the existence of the vapor layer 
is crucial in understanding the spreading
dynamics and heat transfer of droplet impact
on heated surface, 
there has been hardly any experimental 
measurement of the thickness of the vapor layer to date.
Here, we provide direct measurements of the vapor thickness
of drop impact in the gentle film boiling regime
using interferometry. 
In Fig.~\ref{fig:thickness}a,
we show the interference pattern from a bottom view
of an impinging drop at $\We=3.5$ and $T=350^o$C.
The novelty here is that by using 
a color high-speed camera (Photron SA2),
we are able to simultaneously obtain interference patterns
formed by light of different wavelengths (Fig.~\ref{fig:thickness}a). 
The fringe spacings for different lights
are then used to extract the absolute thickness of the vapor
\cite{roeland11}.   
In Fig.~\ref{fig:thickness}b, we show the
measured vapor layer profile. Even for the drop at 
this low Weber number,
the vapor thickness is one order of magnitude
smaller than that in the case of a static 
Leidenfrost drop at a similar surface 
temperature (as predicted by Gottfried {\it et al.} \cite{gottfried66}
and verified experimentally 
by Biance {\it et al.} \cite{biance03}, the vapor thickness is roughly
$20\,\mu$m in the static case), 
consistent with our finding that higher velocities require higher 
surface temperature for the gentle film boiling regime 
to occur.

%%%%%Discussion and conclusions
In conclusion, we have experimentally
explored the $(\We,T)$ phase space
of impact of liquid droplets on heated smooth 
surfaces. The impact behavior can be
separated into three regimes: 
contact boiling, gentle film boiling, and
spraying film boiling.
 We show that the transition temperature
from the contact boiling regime to the gentle 
film boiling regime (the dynamic Leidenfrost
temperature $T_L$ ) increases monotonically
with increasing Weber number. 
We also find that the transition temperature from the gentle film boiling 
 to spraying film boiling regime is
related to boiling bubbles inside the liquid film and 
 deceases with increasing $\We$.
For impacting droplets in 
 both the gentle and the spraying film boiling regimes
 (both occurring when the surface temperature is higher than $T_L$),
 the maximum deformation
displays universality regardless of the 
variation in surface temperature and 
liquid's properties. 

%From our experiments, we infer that 

%%%%%%%%%%%%%%%%%%%%%%%%%%%%%%%%%
%\begin{acknowledgments}
This study was financially supported by the 
European Research Council ERC.
%\end{acknowledgments}

%\bibliography{impact_phase_diagram.bib}

\begin{thebibliography}{19}%
\makeatletter
\providecommand \@ifxundefined [1]{%
 \@ifx{#1\undefined}
}%
\providecommand \@ifnum [1]{%
 \ifnum #1\expandafter \@firstoftwo
 \else \expandafter \@secondoftwo
 \fi
}%
\providecommand \@ifx [1]{%
 \ifx #1\expandafter \@firstoftwo
 \else \expandafter \@secondoftwo
 \fi
}%
\providecommand \natexlab [1]{#1}%
\providecommand \enquote  [1]{``#1''}%
\providecommand \bibnamefont  [1]{#1}%
\providecommand \bibfnamefont [1]{#1}%
\providecommand \citenamefont [1]{#1}%
\providecommand \href@noop [0]{\@secondoftwo}%
\providecommand \href [0]{\begingroup \@sanitize@url \@href}%
\providecommand \@href[1]{\@@startlink{#1}\@@href}%
\providecommand \@@href[1]{\endgroup#1\@@endlink}%
\providecommand \@sanitize@url [0]{\catcode `\\12\catcode `\$12\catcode
  `\&12\catcode `\#12\catcode `\^12\catcode `\_12\catcode `\%12\relax}%
\providecommand \@@startlink[1]{}%
\providecommand \@@endlink[0]{}%
\providecommand \url  [0]{\begingroup\@sanitize@url \@url }%
\providecommand \@url [1]{\endgroup\@href {#1}{\urlprefix }}%
\providecommand \urlprefix  [0]{URL }%
\providecommand \Eprint [0]{\href }%
\@ifxundefined \urlstyle {%
  \providecommand \doi  [0]{\begingroup \@sanitize@url \@doi}%
  \providecommand \@doi [1]{\endgroup \@@startlink {\doibase
  #1}doi:\discretionary {}{}{}#1\@@endlink }%
}{%
  \providecommand \doi  [0]{doi:\discretionary{}{}{}\begingroup
  \urlstyle{rm}\Url }%
}%
\providecommand \doibase [0]{http://dx.doi.org/}%
\providecommand \Doi [0]{\begingroup \@sanitize@url \@Doi }%
\providecommand \@Doi  [1]{\endgroup\@@startlink{\doibase#1}\@@Doi}%
\providecommand \@@Doi [1]{#1\@@endlink}%
\providecommand \selectlanguage [0]{\@gobble}%
\providecommand \bibinfo  [0]{\@secondoftwo}%
\providecommand \bibfield  [0]{\@secondoftwo}%
\providecommand \translation [1]{[#1]}%
\providecommand \BibitemOpen [0]{}%
\providecommand \bibitemStop [0]{}%
\providecommand \bibitemNoStop [0]{.\EOS\space}%
\providecommand \EOS [0]{\spacefactor3000\relax}%
\providecommand \BibitemShut  [1]{\csname bibitem#1\endcsname}%
%</preamble>
\bibitem [{\citenamefont {Leidenfrost}(1756)}]{leiden1756}%
  \BibitemOpen
  \bibfield  {author} {\bibinfo {author} {\bibfnamefont {J.}~\bibnamefont
  {Leidenfrost}},\ }\href@noop {} {\bibfield  {journal} {\bibinfo  {journal}
  {Duisburg}} (\bibinfo {year} {1756})},\ \bibinfo {note} {{\rm T}ranslation:
  On the fixation of water in diverse fire, Int. J. Heat Mass Transfer {\bf 9},
  1153 (1966)}\BibitemShut {NoStop}%
\bibitem [{\citenamefont {Bernardin}\ and\ \citenamefont
  {Mudawar}(1999)}]{bernardin99}%
  \BibitemOpen
  \bibfield  {author} {\bibinfo {author} {\bibfnamefont {J.}~\bibnamefont
  {Bernardin}}\ and\ \bibinfo {author} {\bibfnamefont {I.}~\bibnamefont
  {Mudawar}},\ }\href@noop {} {\bibfield  {journal} {\bibinfo  {journal} {J.
  Heat Transfer},\ }\textbf {\bibinfo {volume} {121}},\ \bibinfo {pages} {894}
  (\bibinfo {year} {1999})}\BibitemShut {NoStop}%
\bibitem [{\citenamefont {Gottfried}\ \emph {et~al.}(1966)\citenamefont
  {Gottfried}, \citenamefont {Lee},\ and\ \citenamefont {Bell}}]{gottfried66}%
  \BibitemOpen
  \bibfield  {author} {\bibinfo {author} {\bibfnamefont {B.}~\bibnamefont
  {Gottfried}}, \bibinfo {author} {\bibfnamefont {C.}~\bibnamefont {Lee}}, \
  and\ \bibinfo {author} {\bibfnamefont {K.}~\bibnamefont {Bell}},\ }\href@noop
  {} {\bibfield  {journal} {\bibinfo  {journal} {Int. J. Heat Mass Transfer},\
  }\textbf {\bibinfo {volume} {9}},\ \bibinfo {pages} {1167} (\bibinfo {year}
  {1966})}\BibitemShut {NoStop}%
\bibitem [{\citenamefont {Bernardin}\ \emph {et~al.}(1997)\citenamefont
  {Bernardin}, \citenamefont {Stebbins},\ and\ \citenamefont
  {Mudawar}}]{bernardin97}%
  \BibitemOpen
  \bibfield  {author} {\bibinfo {author} {\bibfnamefont {J.}~\bibnamefont
  {Bernardin}}, \bibinfo {author} {\bibfnamefont {C.}~\bibnamefont {Stebbins}},
  \ and\ \bibinfo {author} {\bibfnamefont {I.}~\bibnamefont {Mudawar}},\
  }\href@noop {} {\bibfield  {journal} {\bibinfo  {journal} {Int. J. Heat Mass
  Transfer},\ }\textbf {\bibinfo {volume} {40}},\ \bibinfo {pages} {247}
  (\bibinfo {year} {1997})}\BibitemShut {NoStop}%
\bibitem [{\citenamefont {Chandra}\ and\ \citenamefont
  {Avedisian}(1991)}]{chandra91}%
  \BibitemOpen
  \bibfield  {author} {\bibinfo {author} {\bibfnamefont {S.}~\bibnamefont
  {Chandra}}\ and\ \bibinfo {author} {\bibfnamefont {C.}~\bibnamefont
  {Avedisian}},\ }\href@noop {} {\bibfield  {journal} {\bibinfo  {journal}
  {Proc. R. Soc. London, Ser. A},\ }\textbf {\bibinfo {volume} {432}},\
  \bibinfo {pages} {13} (\bibinfo {year} {1991})}\BibitemShut {NoStop}%
\bibitem [{\citenamefont {Anders}\ \emph {et~al.}(1993)\citenamefont {Anders},
  \citenamefont {Roth},\ and\ \citenamefont {Frohn}}]{anders93}%
  \BibitemOpen
  \bibfield  {author} {\bibinfo {author} {\bibfnamefont {K.}~\bibnamefont
  {Anders}}, \bibinfo {author} {\bibfnamefont {N.}~\bibnamefont {Roth}}, \ and\
  \bibinfo {author} {\bibfnamefont {A.}~\bibnamefont {Frohn}},\ }\href@noop {}
  {\bibfield  {journal} {\bibinfo  {journal} {Exp. Fluids},\ }\textbf {\bibinfo
  {volume} {15}},\ \bibinfo {pages} {91} (\bibinfo {year} {1993})}\BibitemShut
  {NoStop}%
\bibitem [{\citenamefont {Xiong}\ and\ \citenamefont {Yuen}(1991)}]{xiong91}%
  \BibitemOpen
  \bibfield  {author} {\bibinfo {author} {\bibfnamefont {T.}~\bibnamefont
  {Xiong}}\ and\ \bibinfo {author} {\bibfnamefont {M.}~\bibnamefont {Yuen}},\
  }\href@noop {} {\bibfield  {journal} {\bibinfo  {journal} {Int. J. Heat Mass
  Transfer},\ }\textbf {\bibinfo {volume} {34}},\ \bibinfo {pages} {1881}
  (\bibinfo {year} {1991})}\BibitemShut {NoStop}%
\bibitem [{\citenamefont {Moreira}\ \emph {et~al.}(2010)\citenamefont
  {Moreira}, \citenamefont {Moita},\ and\ \citenamefont
  {Pan{\~a}o}}]{moreira10}%
  \BibitemOpen
  \bibfield  {author} {\bibinfo {author} {\bibfnamefont {A.}~\bibnamefont
  {Moreira}}, \bibinfo {author} {\bibfnamefont {A.}~\bibnamefont {Moita}}, \
  and\ \bibinfo {author} {\bibfnamefont {M.}~\bibnamefont {Pan{\~a}o}},\
  }\href@noop {} {\bibfield  {journal} {\bibinfo  {journal} {Prog. Energy
  Combust. Sci.},\ }\textbf {\bibinfo {volume} {36}},\ \bibinfo {pages} {554}
  (\bibinfo {year} {2010})}\BibitemShut {NoStop}%
\bibitem [{\citenamefont {Rein}(2002)}]{rein02}%
  \BibitemOpen
  \bibfield  {author} {\bibinfo {author} {\bibfnamefont {M.}~\bibnamefont
  {Rein}},\ }\href@noop {} {\emph {\bibinfo {title} {Drop-surface
  interactions}}},\ Vol.\ \bibinfo {volume} {428}\ (\bibinfo  {publisher}
  {Springer Verlag Wien},\ \bibinfo {year} {2002})\BibitemShut {NoStop}%
\bibitem [{\citenamefont {Yao}\ and\ \citenamefont {Cai}(1988)}]{yao88}%
  \BibitemOpen
  \bibfield  {author} {\bibinfo {author} {\bibfnamefont {S.}~\bibnamefont
  {Yao}}\ and\ \bibinfo {author} {\bibfnamefont {K.}~\bibnamefont {Cai}},\
  }\href@noop {} {\bibfield  {journal} {\bibinfo  {journal} {Exp. Therm Fluid
  Sci.},\ }\textbf {\bibinfo {volume} {1}},\ \bibinfo {pages} {363} (\bibinfo
  {year} {1988})}\BibitemShut {NoStop}%
\bibitem [{\citenamefont {Bernardin}\ and\ \citenamefont
  {Mudawar}(2004)}]{bernardin04}%
  \BibitemOpen
  \bibfield  {author} {\bibinfo {author} {\bibfnamefont {J.}~\bibnamefont
  {Bernardin}}\ and\ \bibinfo {author} {\bibfnamefont {I.}~\bibnamefont
  {Mudawar}},\ }\href@noop {} {\bibfield  {journal} {\bibinfo  {journal} {J.
  Heat Transfer},\ }\textbf {\bibinfo {volume} {126}},\ \bibinfo {pages} {272}
  (\bibinfo {year} {2004})}\BibitemShut {NoStop}%
\bibitem [{\citenamefont {Wang}\ \emph {et~al.}(2000)\citenamefont {Wang},
  \citenamefont {Lin},\ and\ \citenamefont {Chen}}]{wang00}%
  \BibitemOpen
  \bibfield  {author} {\bibinfo {author} {\bibfnamefont {A.}~\bibnamefont
  {Wang}}, \bibinfo {author} {\bibfnamefont {C.}~\bibnamefont {Lin}}, \ and\
  \bibinfo {author} {\bibfnamefont {C.}~\bibnamefont {Chen}},\ }\href@noop {}
  {\bibfield  {journal} {\bibinfo  {journal} {Phys. Fluids},\ }\textbf
  {\bibinfo {volume} {12}},\ \bibinfo {pages} {1622} (\bibinfo {year}
  {2000})}\BibitemShut {NoStop}%
\bibitem [{\citenamefont {Celata}\ \emph {et~al.}(2006)\citenamefont {Celata},
  \citenamefont {Cumo}, \citenamefont {Mariani},\ and\ \citenamefont
  {Zummo}}]{celata06}%
  \BibitemOpen
  \bibfield  {author} {\bibinfo {author} {\bibfnamefont {G.}~\bibnamefont
  {Celata}}, \bibinfo {author} {\bibfnamefont {M.}~\bibnamefont {Cumo}},
  \bibinfo {author} {\bibfnamefont {A.}~\bibnamefont {Mariani}}, \ and\
  \bibinfo {author} {\bibfnamefont {G.}~\bibnamefont {Zummo}},\ }\href@noop {}
  {\bibfield  {journal} {\bibinfo  {journal} {Heat Mass Transfer},\ }\textbf
  {\bibinfo {volume} {42}},\ \bibinfo {pages} {885} (\bibinfo {year}
  {2006})}\BibitemShut {NoStop}%
\bibitem [{par()}]{particles}%
  \BibitemOpen
  \href@noop {} {}\bibinfo {note} {A systematic study of the effect of
  particles on the transition temperature is beyond the scope of this
  paper.}\BibitemShut {Stop}%
\bibitem [{\citenamefont {Tsai}\ \emph {et~al.}(2011)\citenamefont {Tsai},
  \citenamefont {Hendrix}, \citenamefont {Dijkstra}, \citenamefont {Shui},\
  and\ \citenamefont {Lohse}}]{tsai11}%
  \BibitemOpen
  \bibfield  {author} {\bibinfo {author} {\bibfnamefont {P.}~\bibnamefont
  {Tsai}}, \bibinfo {author} {\bibfnamefont {M.}~\bibnamefont {Hendrix}},
  \bibinfo {author} {\bibfnamefont {R.}~\bibnamefont {Dijkstra}}, \bibinfo
  {author} {\bibfnamefont {L.}~\bibnamefont {Shui}}, \ and\ \bibinfo {author}
  {\bibfnamefont {D.}~\bibnamefont {Lohse}},\ }\Doi {10.1039/C1SM05801K}
  {\bibfield  {journal} {\bibinfo  {journal} {Soft Matter},\ } (\bibinfo {year}
  {2011})},\ \bibinfo {note} {{\rm DOI}: 10.1039/c1sm05801k}\BibitemShut
  {NoStop}%
\bibitem [{\citenamefont {Chaves}\ \emph {et~al.}(1999)\citenamefont {Chaves},
  \citenamefont {Kubitzek},\ and\ \citenamefont {Obermeier}}]{chaves99}%
  \BibitemOpen
  \bibfield  {author} {\bibinfo {author} {\bibfnamefont {H.}~\bibnamefont
  {Chaves}}, \bibinfo {author} {\bibfnamefont {A.}~\bibnamefont {Kubitzek}}, \
  and\ \bibinfo {author} {\bibfnamefont {F.}~\bibnamefont {Obermeier}},\
  }\href@noop {} {\bibfield  {journal} {\bibinfo  {journal} {Int. J. Heat Fluid
  Flow},\ }\textbf {\bibinfo {volume} {20}},\ \bibinfo {pages} {470} (\bibinfo
  {year} {1999})}\BibitemShut {NoStop}%
\bibitem [{\citenamefont {Clanet}\ \emph {et~al.}(2004)\citenamefont {Clanet},
  \citenamefont {B{\'e}guin}, \citenamefont {Richard},\ and\ \citenamefont
  {Qu{\'e}r{\'e}}}]{clanet04}%
  \BibitemOpen
  \bibfield  {author} {\bibinfo {author} {\bibfnamefont {C.}~\bibnamefont
  {Clanet}}, \bibinfo {author} {\bibfnamefont {C.}~\bibnamefont {B{\'e}guin}},
  \bibinfo {author} {\bibfnamefont {D.}~\bibnamefont {Richard}}, \ and\
  \bibinfo {author} {\bibfnamefont {D.}~\bibnamefont {Qu{\'e}r{\'e}}},\
  }\href@noop {} {\bibfield  {journal} {\bibinfo  {journal} {J. Fluid Mech.},\
  }\textbf {\bibinfo {volume} {517}},\ \bibinfo {pages} {199} (\bibinfo {year}
  {2004})}\BibitemShut {NoStop}%
\bibitem [{\citenamefont {van~der Veen}\ \emph {et~al.}(2011)\citenamefont
  {van~der Veen}, \citenamefont {Tran}, \citenamefont {Lohse},\ and\
  \citenamefont {Sun}}]{roeland11}%
  \BibitemOpen
  \bibfield  {author} {\bibinfo {author} {\bibfnamefont {R.}~\bibnamefont
  {van~der Veen}}, \bibinfo {author} {\bibfnamefont {T.}~\bibnamefont {Tran}},
  \bibinfo {author} {\bibfnamefont {D.}~\bibnamefont {Lohse}}, \ and\ \bibinfo
  {author} {\bibfnamefont {C.}~\bibnamefont {Sun}},\ }\href@noop {} { (\bibinfo
  {year} {2011})},\ \bibinfo {note} {submitted}\BibitemShut {NoStop}%
\bibitem [{\citenamefont {Biance}\ \emph {et~al.}(2003)\citenamefont {Biance},
  \citenamefont {Clanet},\ and\ \citenamefont {Qu{\'e}r{\'e}}}]{biance03}%
  \BibitemOpen
  \bibfield  {author} {\bibinfo {author} {\bibfnamefont {A.}~\bibnamefont
  {Biance}}, \bibinfo {author} {\bibfnamefont {C.}~\bibnamefont {Clanet}}, \
  and\ \bibinfo {author} {\bibfnamefont {D.}~\bibnamefont {Qu{\'e}r{\'e}}},\
  }\href@noop {} {\bibfield  {journal} {\bibinfo  {journal} {Phys. Fluids},\
  }\textbf {\bibinfo {volume} {15}},\ \bibinfo {pages} {1632} (\bibinfo {year}
  {2003})}\BibitemShut {NoStop}%
\end{thebibliography}

%merlin.mbs 2010-03-15 4.21a (PWD, AO, DPC)
%Control: key (0)
%Control: author (8) initials jnrlst
%Control: editor formatted (1) identically to author
%Control: production of article title (-1) disabled
%Control: page (0) single
%Control: year (1) truncated
%Control: production of eprint (0) enabled
%

\end{document}